\documentclass[a4paper,12pt]{article}
\usepackage[main=english]{babel}
\usepackage[utf8]{inputenc}
\usepackage[T1]{fontenc}
\usepackage{lmodern}
\usepackage{cite}
\usepackage{mathrsfs}
\usepackage{latexsym,amssymb,amscd,amsthm,amsmath}
\usepackage{mathtools}

\newtheorem{definition}{Definição}[section]

\usepackage{mathrsfs}

\title{An Approach to the Primordial Universe Using Colombeau's Simplified Algebra}
\author{Jonatas A. Silva \and Fábio C. Carvalho \and Antonio R. G. Garcia}

\begin{document}
\maketitle

\begin{abstract}
The proposal no {\it{boundary}} of physicists Hartle and Hawking seeks
to build a satisfactory model of the early Universe, in a way that
avoids the singularity ({\it{Big Bang}}) of the beginning of the Universe.
As a consequence of this proposal, the concept of metric signature
change arises, which is approached in different ways in the literature.
Here, we reinterpret the Mansouri-Nozari approach, which modifies the FLRW
metric, and uses the formalism of Colombeau’s Algebras, to develop its equations.
In addition, we write the function that changes the sign in terms of redshift.
Finally, we developed Friedmann’s equations, of the modified metric,
as well as the equation of state and other relevant equations in Cosmology.
\end{abstract}

\section*{Introduction}
The idea of changing the signature had its origins in Hartle-Hawking's no-boundary proposal \cite{Hartle1983,Hawking1984}. This proposal investigates the wave function of the Universe and seeks to build a satisfactory model of the Universe in order to avoid the primordial spacetime singularity predicted by the standard cosmological model, using a combination of general relativity and quantum mechanics. One of the intriguing features of this proposal is the idea that the spacetime signature must change in very primitive times, resulting in an origin of the Universe in a regime where there is no time, so that ``spacetime'' was initially Euclidean and, by changing the signature of the ``space-time'' metric, the transition to the usual Lorentzian space-time occurred \cite{Ellis1992,Mansouri1999, Nozari2006}.

To investigate the consequences of such a situation, we use the Friedmann-Lemaître-Robertson-Walker (FLRW) metric modified by a function $f(t)$, responsible for the signature change, following the works \cite{Mansouri1999,Nozari2006}. Since Einstein's equations of general relativity are nonlinear PDEs, nonlinear operations between distributions in discontinuous metrics are unavoidable. In this sense, we make use of Colombeau's algebras to skirt this difficulty.

Colombeau's theory proved to be very consistent with applications to physical problems. Cosmic strings \cite{Clarke1996}, Reissner–Nordström metric \cite{Steinbauer1997}, general relativity \cite{Vickers2012, Grosser2001}. In addition, there is a vast production of articles related to problems intrinsic to mathematics, as well as the development of basic concepts, which makes the formalism even more solid. Generalized numbers \cite{Aragona2013}, generalized solutions of nonlinear parabolic equations \cite{Aragona2009}, generalized quaternions \cite{Cortes2017}, off-diagonal condition \cite{Rodrigues2022}. These are just some of the applications of the formalism of the Colombeau algebras, as well as being stopping points to appreciate and understand the theory and the wide spectrum of applications.

In this paper, as announced we use Colombeau generalized functions to deal with the nonlinear operations that arise in the development of FLRW solutions of Einstein's equations. Colombeau's algebras \cite{Aragona1991,Colombeau1984} are differential algebras, commutative, associative, distributive of multiplication with respect to addition, and worth Leibniz's rule for the derivative of the product of functions, and contain the distributions by continuous immersion.

In this article, we develop a cosmological model from the FLRW metric modified by \cite{Mansouri1999, Nozari2006}, in the context of the signature change. We reinterpret the function $f(t)$ responsible for the sign change and express it in terms of the redshift. With this, we develop the Friedmann equations and other relevant expressions in cosmology.



This paper is organized as follows. Section \ref{preli}, presents an overview of the fundamentals of Colombeau algebra. In Section 2, we develop our proposal. Section 3, presents the development of Friedmann's equations, the conservation equation and the state equation. In Section 4, some cosmological parameters were developed. Summary and conclusions are provided in Section 5. We use the signature $(- +~+~+)$ for Lorentzian manifolds and follow the curvature conventions of Misner et al. \cite{Misner}.

\section{Preliminary}\label{preli}

Colombeau algebras are differential, commutative, associative, distributive algebras of multiplication in relation to addition, which holds the Leibniz rule for the derivative of the product of functions and contains distributions by continuous embedding. These algebras are defined as follows.




\begin{definition}\label{colom-1}
Let $\Omega$ be an open subset of $\mathbb{R}^n$. We define $\mathcal{E}(\Omega)\coloneqq \mathscr{C}^\infty(\Omega)^I$ which is a point-to-point operation ring and hence the subring
	\begin{eqnarray}
	\mathcal{E}_M(\Omega)\coloneqq\{(u_\varepsilon)_\varepsilon\in\mathcal{E}(\Omega)|\forall~K\subset\subset\Omega,
	~\forall~\alpha\in\mathbb{N}_0^n,~\exists~p\in\mathbb{N}&&\nonumber\\
	\mbox{with}~\sup_{x\in
		K}|\partial^\alpha
	u_\varepsilon(x)|=O(\varepsilon^{-p})~\mbox{as}~ \varepsilon\to
	0\}
	\end{eqnarray}
and its maximal ideal
	\begin{eqnarray}
	\mathcal{N}(\Omega)\coloneqq\{(u_\varepsilon)_\varepsilon\in\mathcal{E}_M(\Omega)|\forall~K\subset\subset\Omega,
	~\forall~\alpha\in\mathbb{N}_0^n
	~\mbox{e}~\forall~q\in\mathbb{N}&&\nonumber\\
	\sup_{x\in K}|\partial^\alpha
	u_\varepsilon(x)|=O(\varepsilon^q)~\mbox{as}~ \varepsilon\to
	0\}.
	\end{eqnarray}
	The Colombeau algebra ``simplified'' $\mathscr{G}(\Omega)$ is defined as the quotient space \[\mathscr{G}(\Omega)\coloneqq\mathcal{E}_M(\Omega)/\mathcal{N}(\Omega).\]
\end{definition}

By Definition \ref{colom-1}, we observe that a simplified generalized Colombeau function is therefore a family of moderate functions $(f_\varepsilon(\cdot))\in\mathscr{C}^\infty$ modulo as null functions. 

In order to define the product of two given distributions the route taken here is to substitute for one or both factors a smooth regularization (obtained by convolution with a so-called mollifier), compute the product in $\mathscr{C}^\infty\times\mathscr{D}^\prime$ or $\mathscr{C}^\infty\times\mathscr{C}^\infty$, and then pass to the limit, if possible.

\begin{definition}\label{produto}
For $u, v \in \mathscr{D}^\prime$ set
	\item[$i)$] $u\cdot [v] = \displaystyle\lim_{\varepsilon\rightarrow 0}u(v\ast \rho_{\varepsilon});$
	\item[$ii)$] $[u]\cdot v = \displaystyle\lim_{\varepsilon\rightarrow 0}(u\ast \rho_{\varepsilon})v;$
	\item[$iii)$] $[u]\cdot [v] = \displaystyle\lim_{\varepsilon\rightarrow 0}(u\ast \rho_{\varepsilon})(v\ast \sigma_{\varepsilon});$
	\item[$iv)$] $[u\cdot v] = \displaystyle\lim_{\varepsilon\rightarrow 0}(u\ast \rho_{\varepsilon})(v\ast \rho_{\varepsilon})$ \\	
	if the limit exists in $\mathscr{D}^\prime$ for all strict delta nets $(\rho_{\varepsilon})_{\varepsilon}$ or $(\rho_{\varepsilon})_{\varepsilon}$ and $(\sigma_{\varepsilon})_{\varepsilon}$, respectively. 
\end{definition} 
\noindent
The test functions are $\varphi\in\mathscr{C}_0^\infty$ where the regularizing net is defined $\varphi_\varepsilon(x)=\varepsilon^{-n}\varphi\left(\frac{x}{\varepsilon}\right),~\forall~\varepsilon\in ]0,1]$.
Definition \ref{produto} is of great importance in our work because it is through it that we understand the potential and products of distribution.




The construction of the space of the simplified generalized Colombeau functions, $\mathscr{G}(\Omega)$, was organized so as to obtain an immersion by means of the convolution operation with a suitable mollifier. Now, from what we saw earlier, this will require that such a mollifier satisfies the following properties:

\begin{enumerate}
	\item[$i)$] $\displaystyle\int_{\mathbb{R}^n}\rho(x)dx=1$;
	\item[$ii)$] $\displaystyle\int_{\mathbb{R}^n} x^\alpha\rho(x)dx=0, ~\forall~|\alpha|\ge 1$.
\end{enumerate}

Due to the construction of the algebras of generalized Colombeau functions, we have that given $u\in\mathscr{D}^\prime(\Omega)$, the net of functions $u_\varepsilon=(u\ast\varphi_\varepsilon)\in\mathscr{G}(\Omega),~\forall~\varepsilon\in]0,1]$, where, as we defined earlier, for $\varphi\in\mathscr{D}(\Omega)$, we obtain its regularizer $\varphi_\varepsilon(x)=\varepsilon^{-n}\varphi\left(\frac{x}{\varepsilon}\right)$. And thus we have that $\mathscr{D}^\prime(\Omega)\hookrightarrow\mathscr{G}(\Omega)$. This section was structured and presented as \cite{Kunzinger1996}.

\section{Change of signature of the metric}

The signature change addressed by \cite{Mansouri1999} has as its starting point the modified FLRW metric, which is given by
\begin{eqnarray}\label{Eq 4.2}
ds^2 = -f(t)dt^2 + a^2(t)\left[\dfrac{dr^2}{1-kr^2} + r^2d\theta^2 + r^2\sin^2\theta d\phi\right],
\end{eqnarray}
where the function $f(t)$ is responsible for changing the signature of the metric and is defined by
\begin{eqnarray}\label{Eq 4.3}
f(t)=\theta(t)-\theta(-t)
\end{eqnarray}
and $\theta(t)$ is a ``Heaviside-like'' function, it will be so called, as it is different from how the Heaviside function (or step function) is normally defined. Furthermore it is considered that $a^2(t) = a^2_{+}(t)\theta(t) - a^2_{-}(t)\theta(-t).$ The function $\theta (t)$ is given by
\begin{equation}\label{Eq 4.4}
\theta(t) = \left\{
\begin{array}{ccc}
1 &\textrm{if}& t>0 \\
\tau &\textrm{if}& t=0 \\
0 &\textrm{if}& t<0 \\
\end{array}
\right.,
\end{equation}
where $\tau>\frac{1}{2}$. It is important to note that $\tau$ should not be confused with the notation traditionally adopted in the literature for the proper time. Knowing this, we have to
\begin{equation}\label{Eq 4.5}
\theta(-t) = \left\{
\begin{array}{ccc}
1 &\textrm{if}& t<0 \\
\tau &\textrm{if}& t=0 \\
0 &\textrm{if}& t>0 \\
\end{array}
\right.,
\end{equation}
and satisfies the following property $\theta(-t) = 1 - \theta(t)$. With this, we have that the function $f(t)$ is given by
\begin{equation}\label{Eq 4.6}
f(t) = \left\{
\begin{array}{ccc}
1 &\textrm{if}& t>0 \\
2\tau -1 &\textrm{if}& t=0 \\
-1 &\textrm{if}& t<0 \\
\end{array}
\right..
\end{equation}
Due to the abrupt jump at $t=0$, the regularized function $f_{\varepsilon}$ given through convolution was considered.


In our approach, we consider that the scale factor $a(t)$ is written and has the same characteristics as the standard model, so we do not need to write as proposed by Mansouri-Nozari, we chose this because the transition of the sign change in our proposal takes place in a different way than Mansouri-Nozari, this will become clearer below.

In the Mansouri-Nozari approach we can observe that the function $f(t)$ changes the signature of the metric only when $t<0$, which physically does not make sense. Thinking about it, we reinterpreted the function so that it was physically plausible.

\subsection{Reinterpretation of function $f(t)$.} 

The function $\theta(t)$ can assume any value $c$, such that $0<c<1$, in $t=0$ \cite{RamKanwal1983}. Here, we choose $0<c<\frac{1}{2}$. Therefore, our function $f(t)$ is
\begin{equation}\label{Eq 4.7}
f(t) = \left\{
\begin{array}{ccc}
1 &\textrm{if}& t>0 \\
2c -1 &\textrm{if}& t=0 \\
-1 &\textrm{if}& t<0 \\
\end{array}
\right..
\end{equation}
Where the curve touches the vertical axis depends on the value $c$ taken. By convolution, we can show that, no matter how small $t$, that is, $\varepsilon \rightarrow 0$, the regularized function of $f(t)$ converges to $2c-1$, that is, it changes signal without going through the origin. For this consider the following: from the definition of convolution, we have

\begin{equation}\label{Eq 4.7.a}
(f\ast \varphi_{\varepsilon})(t) = \displaystyle\int_{-\varepsilon}^{+\varepsilon}f(t-y)\varphi_{\varepsilon}(y)dy,
\end{equation}
where $\varphi_\varepsilon(x)=\varepsilon^{-1}\varphi\left(\frac{x}{\varepsilon}\right),~\forall~\varepsilon\in ]0,1]$ is constructed from $\varphi \in \mathscr{C}_0^\infty(\Omega),~ \Omega \subset \mathbb{R},$ from the test function $\varphi$. Therefore, we have to
\begin{eqnarray}\label{Eq 4.7.b}
(f\ast\varphi_{\varepsilon})(t)&=&\displaystyle\int_{-\varepsilon}^{+\varepsilon}f(t-y)\varepsilon^{-1}\varphi\left(\frac{y}{\varepsilon}\right)dy\nonumber\\
&=&\varepsilon^{-1}\displaystyle\int_{-\varepsilon}^{+\varepsilon}f(t)\varphi\left(\frac{y}{\varepsilon}\right)dy
\end{eqnarray}
For the case where $\varepsilon\rightarrow 0$, we have
\begin{eqnarray}\label{Eq 4.7.c}
(f\ast\varphi_{\varepsilon})(t)&=&\varepsilon^{-1}\displaystyle\int_{-\varepsilon}^{+\varepsilon}(2c-1)\varphi\left(\frac{y}{\varepsilon}\right)dy
\end{eqnarray}
Making a change of variable $u=y/\varepsilon$, we have
\begin{eqnarray}\label{Eq 4.7.d}
(f\ast\varphi_{\varepsilon})(t)&=&(2c-1)\displaystyle\int_{-\varepsilon}^{+\varepsilon}\varphi(u)du\nonumber \\
&=&2c-1,
\end{eqnarray}
where $\displaystyle\int_{-\varepsilon}^{+\varepsilon}\varphi(u)du = 1$.

Note that in the vicinity of zero, the convolution of $f(t)$ is $2c-1$, and outside the vicinity is $-1$ or $+1$, this is the correct reading of the convolution of f(t). Also, note that in the \textit{Big Bang} ($t=0$) the convolution of $f(t)$ is worth $2c-1$, this describes that the transition of the signature change does not occur on the surface $t=0$ but after a sufficiently small interval $0\leq t\leq \varepsilon$, where $\varepsilon\in ]0,1]$, after this value of F the convolution assumes the value $+1$ and we have the usual metric of the standard model.

The condition $c<\frac{1}{2}$ is taken so that the function and its powers do not cancel at any point, making possible operations such as $\dfrac{1}{f}$ and $\dfrac{1}{f^2}$. Later, we will come across operations of type $\dfrac{\dot{f}}{f(t)}~\textrm{and}~\dfrac{\dot{f}}{f^2(t)}$. According to the standard calculation of distributions we have

\begin{eqnarray}
\dot{f}(t) &=& \dot{\theta}(t) - \dot{\theta}(-t)\nonumber\\
&=& \delta(t) - \delta(-t)(-1)\nonumber\\
&=& \delta(t) + \delta(t)\nonumber\\
&=& 2 \delta(t),
\end{eqnarray}
since $\delta(-t) = \delta(t).$ The multiplication of the distribution $\delta(t)$ with the generalized functions $\dfrac{1}{f(t)}~\textrm{e}~\dfrac{1}{f^2(t)}$ is not defined in the Theory of Distributions but is well defined in the sense of Colombeau algebras, as defined in Definition \ref{produto}.

Thus, with the regularization of the function $f(t)$ it is possible to have a change of signature in the FLRW metric in a period that $t>0$. Making possible a study in the interval in which the space-time had a Euclidean region, i.e., positive signature.

According to \cite{Ellis1992}, this change may have occurred at the Planck epoch in the FLRW universe, or at a less extreme period, such as the GUTs epoch. Regardless of the stage, the way in which we construct the function $f(t)$ allows for a freedom in when the signature change may have occurred. To do this, simply assign different values to $\varepsilon,~\textrm{onde}~  \varepsilon\in\ ]0,1]$, similarly there is a freedom for $f(t)$ to take on different values at t=0 by simply choosing a different value for $c$, $0<c<\frac{1}{2}$.

\begin{figure}[htbp!] 
	\centering
	\label{ft2}
	\includegraphics[scale=0.3]{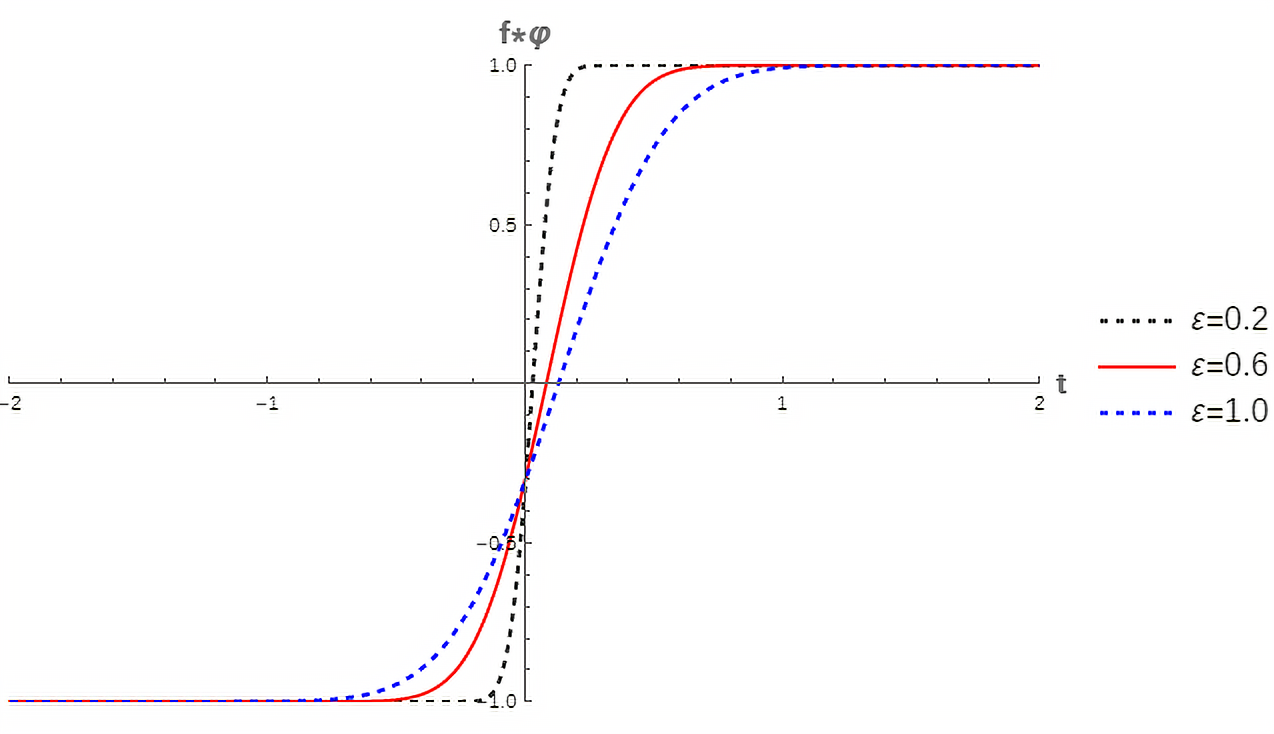} \\
	\caption{Function $f(t)$ regularized by convolution.}
\end{figure}

For the sake of analysis, without worrying about which period the transition occurred, let's look at the convolution plot of $f(t)$ for different values of $\varepsilon$ and $c=0,35$, that is, at the origin the function assumes the value $-0,3$ (see Fig. \ref{ft2}).

It is notorious that there is a smooth evolution of the function in a period $t>0$, as well as the signal transition. This is an important point for our approach, since the smoothing proposed by \cite{Mansouri1999, Nozari2006} occurs in $t<0$, and does not allow for a satisfactory physical description.
\newpage

\subsection{The function $f(t)$ in terms of redshift.} 

It is convenient to represent function $f$ in terms of redshift because of its importance in Cosmology. For this, we will use the relationship between the cosmic time $t$ and the redshit $z$ given by \cite{carmeli2006}. The conditions for the validity of the relation expressed in (\ref{Eq 4.8}) are established in the previous reference. According to \cite{carmeli2006}, we have

\begin{equation}\label{Eq 4.8}
t(z) = \dfrac{2H_{0}^{-1}}{1 + (1+z)^2}.
\end{equation}
From that, we get
\begin{equation}\label{Eq 4.9}
z(t) = \sqrt{\dfrac{2H_{0}^{-1}}{t}-1} -1.
\end{equation}
Note that $t\circ z(t) = t(z(t)) = t$ e $z\circ t(z) = z(t(z))=z$.

Thus, the function $z(t)$ is the inverse of $t(z)$. Therefore,
\begin{equation}\label{Eq 4.10}
t^{-1}(z) = \sqrt{\dfrac{2H_{0}^{-1}}{z}-1} -1,
\end{equation}
where $0<z<H_{0}^{-1}$ and $z\leq 2H_{0}^{-1}$. With this, we can now define the functions $\theta(t^{-1})$ and $\theta(-t^{-1})$. Using Eq.(\ref{Eq 4.10}) , we have
\begin{equation}\label{Eq 4.11}
\theta(t^{-1}) \equiv \gamma(z) = \left\{
\begin{array}{ccc}
1 &\textrm{se}& \sqrt{\dfrac{2H_{0}^{-1}}{z}-1} -1>0 \Rightarrow z<H_{0}^{-1}  \\
0 &\textrm{se}& \sqrt{\dfrac{2H_{0}^{-1}}{z}-1}-1=0 \Rightarrow z = H_{0}^{-1} \\
c &\textrm{se}& \sqrt{\dfrac{2H_{0}^{-1}}{z}-1} -1<0 \Rightarrow z>H_{0}^{-1}\\
\end{array}
\right..
\end{equation}
Similarly, we obtain
\begin{equation}\label{Eq 4.12}
\gamma(-z) = \left\{
\begin{array}{ccc}
1 &\textrm{if}& z > H_{0}^{-1}  \\
0 &\textrm{if}& z = H_{0}^{-1} \\
c &\textrm{if}& z < H_{0}^{-1}\\
\end{array}
\right..
\end{equation}

Having in hand the functions $\gamma(z)$ and $\gamma(-z)$, we will define a new function $g$, which is given by
\begin{equation}\label{Eq 4.13}
g(z) = \gamma(z) - \gamma(-z),
\end{equation}
that is,
\begin{equation}\label{Eq 4.14}
g(z) = \left\{
\begin{array}{ccc}
1 &\textrm{if}& z < H_{0}^{-1}  \\
2c-1 &\textrm{if}& z = H_{0}^{-1} \\
-1 &\textrm{if}& z > H_{0}^{-1}\\
\end{array}
\right..
\end{equation}
\begin{figure}[htbp!] 
	\centering
	\label{fz}
	\includegraphics[scale=0.7]{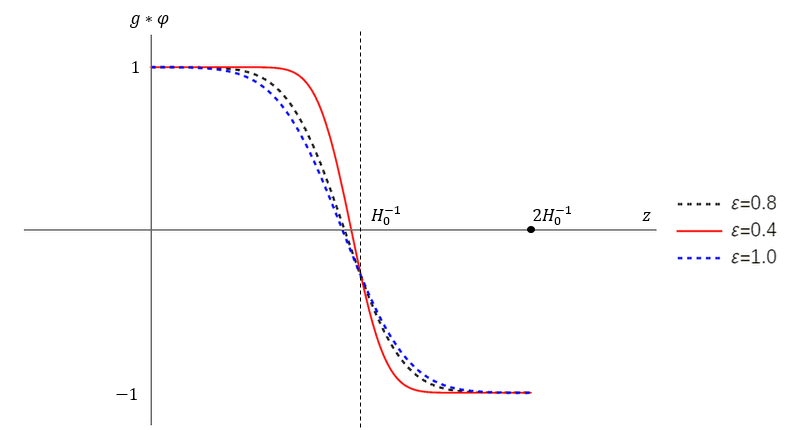} \\
	\caption{Function $g(z)$ regularized by convolution.}
\end{figure}
Analogous to what we did for $f(t)$ we can do for $g(z)$, that is, for a sufficiently small value of $\varepsilon$ the function $g(z)$ converges to $2c-1$, that is, the function does not cancel at the origin and does not cause problems at that point. This way it is possible to analyze the change of the signature of the metric with a new perspective.

In view of the relation of cosmic time $t$ to redshift $z$ expressed by Eq.(\ref{Eq 4.8}) we define a new function $g(z)$. With this it was possible to analyze the signature change from the perspective of redshift. Similarly to the function $f(t)$, we can adjust the value of  $\varepsilon$ so that the change occurs in a different $z$. Let’s see the graph of the convolution of $g(z)$ in Fig. \ref{fz}.

With the regularization of these functions by convolution, they belong to the space of generalized Colombeau functions, and due to the properties of $\varphi_{\varepsilon}$, they become $\mathscr{C}_0^\infty(\mathbb{R})$ functions.
\newpage

\section{Friedmann equations}

From the modified FLRW metric, given by Eq.(\ref{Eq 4.2}), we can deduce the Friedmann equations, and thus study the expansion of the universe in this context. First, we calculate the relevant components of the Einstein tensor for the metric by Eq.(\ref{Eq 4.2})
\begin{eqnarray}\label{Eq 4.15}
G_{00} = \dfrac{3\dot{a}^2}{a^2} - \dfrac{3kf}{a^2}.
\end{eqnarray}
\begin{eqnarray}\label{Eq 4.16}
G_{11} = \dfrac{1}{f(1-kr^2)}\left(-2a\ddot{a} - \dot{a}^2 + kf + \dfrac{\dot{f}}{f}a\dot{a}\right).
\end{eqnarray}
\begin{eqnarray}\label{Eq 4.17}
G_{22} = \dfrac{r^2}{f}\left(-2a\ddot{a} - \dot{a}^2 + kf + \dfrac{\dot{f}}{f}a\dot{a}\right).
\end{eqnarray}
\begin{eqnarray}\label{Eq 4.18}
G_{33} = \dfrac{r^2sen^2\theta}{f}\left(-2a\ddot{a} - \dot{a}^2 + kf + \dfrac{\dot{f}}{f}a\dot{a}\right).
\end{eqnarray}

The energy-momentum tensor of a perfect fluid is given by 
\begin{eqnarray}
T_{\alpha\beta} = (\rho + p)u_{\alpha}u_{\beta} + pg_{\alpha \beta}
\end{eqnarray}

The time component of the energy-momentum tensor is simply the energy density, that is, $T_{00} = \rho$, we obtain the first Friedmann equation
\begin{eqnarray}\label{Eq 4.19}
\left(\dfrac{\dot{a}}{a}\right)^2 - \dfrac{kf}{a^2} = \dfrac{8\pi G \rho f}{3},
\end{eqnarray}
for the case where the Universe is flat, that is, $k=0$, we have
\begin{eqnarray}\label{Eq 4.20}
\left(\dfrac{\dot{a}}{a}\right)^2  = \dfrac{8\pi G \rho f}{3}.
\label{Fri1}
\end{eqnarray}

We also obtain the second Friedmann equation, considering the spatial components of the Einstein equations (remember that the spatial components of the energy-momentum tensor are simply the pressure, that is,$T_{ij} = p, i=j$), that is,
\begin{eqnarray}\label{Eq 4.21.a}
G_{ij} = 8\pi G T_{ij}.
\end{eqnarray}
Substituting equations (\ref{Eq 4.16}) - (\ref{Eq 4.18}), into the previous equation, and using $T_{ij} = p,\textrm{for}~ i=j$, we obtain
\begin{eqnarray}\label{Eq 4.21}
\dfrac{2}{f}\left(\dfrac{\ddot{a}}{a}\right) + \dfrac{1}{f}\left(\dfrac{\dot{a}}{a}\right)^2 - \dfrac{k}{a^2} - \dfrac{\dot{f}\dot{a}}{f^2 a}  = -8\pi G p.
\end{eqnarray}
In the particular case where $k=0$ the equation is given by
\begin{eqnarray}\label{Eq 4.22}
\dfrac{2}{f}\left(\dfrac{\ddot{a}}{a}\right) + \dfrac{1}{f}\left(\dfrac{\dot{a}}{a}\right)^2 - \dfrac{\dot{f}\dot{a}}{f^2 a}  = -8\pi G p.
\label{Fri2}
\end{eqnarray}
Substituting Eq.(\ref{Eq 4.20}) into Eq.(\ref{Eq 4.22}), we have
\begin{eqnarray}\label{Eq 4.23}
\left(\dfrac{\ddot{a}}{a}\right) = -\dfrac{4\pi G}{3}\left[ \rho f + 3pf - \dfrac{\dot{f}}{f}\left(\dfrac{3}{8\pi G}\right)^{\frac{1}{2}} \sqrt{\rho f}\right].
\end{eqnarray}
We can rewrite this equation in a more compact form as follows
\begin{eqnarray}\label{Eq 4.24}
\left(\dfrac{\ddot{a}}{a}\right) = -\dfrac{4\pi G}{3}\left( \rho_{T} + 3P_{T}\right),
\end{eqnarray}
where $\rho_{T} \equiv \rho f - \dfrac{\dot{f}}{f}\left(\dfrac{3}{8\pi G}\right)^{\frac{1}{2}} \sqrt{\rho f}$ e $P_{T} \equiv pf$, which is the total density and the total pressure, written in terms of function F, respectively. Note that the previously obtained equations fall in the standard case when the function is equal to one.

Now, let's express the conservation equation. For this, we will derive Eq.(\ref{Eq 4.20}) with respect to $t$. Thus, we have
\begin{eqnarray}\label{Eq 4.25}
\dfrac{8\pi G}{3} \dot{\rho} = -\dfrac{\dot{f}}{f^2}\left(\dfrac{\dot{a}}{a}\right)^2 + \left(\dfrac{\dot{a}}{a}\right)\left[\dfrac{2\ddot{a}}{f a} - \dfrac{2}{f}\left(\dfrac{\dot{a}}{a}\right)^2 \right].
\end{eqnarray}
Substituting equations (\ref{Eq 4.20}) and (\ref{Eq 4.22}) into Eq.(\ref{Eq 4.25}) and performing some operations, we have
\begin{eqnarray}\label{Eq 4.26}
\dfrac{8\pi G}{3} \dot{\rho} = -8\pi G\dfrac{\dot{a}}{a}(p + \rho).
\end{eqnarray}
Therefore, we obtain the conservation equation
\begin{eqnarray}\label{Eq 4.27}
\dot{\rho} = -3H(p + \rho),
\end{eqnarray}
which describes the evolution of energy density in relation to time, being $H = \dfrac{\dot{a}}{a}$.

The set of equations obtained here describes the dynamics of the Universe. We can evaluate the conditions for the accelerated expansion of the Universe. So for that $\ddot{a}>0$, we must have $\rho_{T} + 3P_{T}<0$, so
\begin{eqnarray}\label{Eq 4.27.A}
P_{T}<-\dfrac{1}{3}\rho_{T}.
\end{eqnarray}
We will evidence this condition in terms of the functions $f$ and $\dot{f}$. From Eq.(\ref{Eq 4.23}), we have
\begin{eqnarray}\label{Eq 4.27.B}
\rho f + 3pf - \dfrac{\dot{f}}{f}\left(\dfrac{3}{8\pi G}\right)^{\frac{1}{2}} \sqrt{\rho f} <0.
\end{eqnarray}
Using $p = w\rho$ in (\ref{Eq 4.27.B}) we obtain
\begin{eqnarray}\label{Eq 4.27.B1}
\rho f(1 + w) - \dfrac{\dot{f}}{f}\left(\dfrac{3}{8\pi G}\right)^{\frac{1}{2}} \sqrt{\rho f} <0.
\end{eqnarray}
Hence, we can rewrite this inequality in terms of parameter $w$, obtaining the following expression
\begin{eqnarray}\label{Eq 4.27.C}
w < \dfrac{\dot{f}}{3}\left( \dfrac{3}{8\pi G f^{3}} \right)^{\frac{1}{2}}\dfrac{1}{\sqrt{\rho}} - \dfrac{1}{3}.
\end{eqnarray}
We can write this expression more compactly using the energy density given by Friedmann's first equation, Eq.(\ref{Eq 4.20}). This allows us to obtain the condition for $w$ in terms of the Hubble parameter. Thus, we have
\begin{eqnarray}\label{Eq 4.27.D}
w < \dfrac{\dot{f}}{3}\left( \dfrac{3}{8\pi G f^{3}} \right)^{\frac{1}{2}}\left( \dfrac{8\pi G f}{3} \right)^{\frac{1}{2}} - \dfrac{1}{3}.
\end{eqnarray}
Finally, we obtain
\begin{eqnarray}\label{Eq 4.27.E}
w < \dfrac{1}{3H}\dfrac{\dot{f}}{f^2}- \dfrac{1}{3}.
\end{eqnarray}
Note that the result is similar to the usual one, differing by extra terms that arise as a result of the change in the metric. Note also that this condition varies over time, as $f$ and $H$ also depend on time.  In the case of $f$ equals $1$, we obtain the already known result.

From the regularization of the functions $f(t)$ and $g(z)$ and the conditions imposed on them, it was possible to construct the equations presented here. Initially, we used specifically the function $f(t)$ and derived the first Friedmann equation and the equation of acceleration influenced by it so that in a primitive universe, $f(t)$ assumes a different value and as time evolves it assumes value $1$ and returns to the standard case.

Moreover, we also observe that the conservation equation (Eq.\ref{Eq 4.27}) remains invariant. This shows that regardless of the period, whether in the Euclidean or Lorentzian regime, the conservation equation remains, that is, it is not influenced by the change of signature.

Another interesting result is that due to the function $f(t)$, new terms appear in the acceleration equation, written more compactly using $\rho_{T}$  and $P_{T}$. Thus, we describe the conditions necessary for the accelerated expansion of the universe, given by Eq.(\ref{Eq 4.27.A}). With this, we evidenced this condition in terms of parameter W of the state equation. In our case, the parameter varies with time and is written in terms of $f(t)$ and the Hubble parameter, and it is possible to estimate the condition for $w$ in a Euclidean regime (when $t$ is small enough).

\subsection{Equation of state}

The parameter $w$ of the state equation provides the relationship between energy density and pressure as follows \cite{Wang2016}
\begin{eqnarray}\label{Eq 4.28}
w_{i}=\dfrac{p_{i}}{\rho_{i}}.
\end{eqnarray}
This dimensionless parameter can be used in the FLRW equations to define the evolution of an isotropic universe filled with a perfect fluid \cite{Abdul2019}. The $w$ parameter can be a constant or time-dependent function \cite{Tripathi2017}.

Here, let's consider $w=w(t)$ and determine an expression in terms of the Hubble parameter $H$ and the function $g(z)$. Substituting Eq.(\ref{Eq 4.28}) in Eq.(\ref{Eq 4.27}), we have
\begin{eqnarray}\label{Eq 4.29}
\dot{\rho} = -3H\rho(1 + w),
\end{eqnarray} 
we can rewrite it as 
\begin{eqnarray}\label{Eq 4.30}
\dfrac{1}{\rho}\dfrac{d\rho}{dt} = -\dfrac{3}{a}\dfrac{da}{dt}(1 + w).
\end{eqnarray}

Integrating with respect to time, we have
\begin{eqnarray}\label{Eq 4.31}
\int_{\rho_{0}}^{\rho '}\dfrac{d\rho}{\rho} = -3\int_{a_{0}}^{a '}\dfrac{da}{a}(1 + w),
\end{eqnarray}
where $\rho '$ and $a '$ represent arbitrary values of the energy density and the scale factor, respectively. Therefore, it follows that 
\begin{eqnarray}\label{Eq 4.32}
\ln\left(\dfrac{\rho}{\rho_{0}} \right) = -3\int_{a_{0}}^{a '}\dfrac{da}{a}(1 + w).
\end{eqnarray}
With this, we have that 
\begin{eqnarray}\label{Eq 4.33}
\xi = \exp\left(-3\int_{a_{0}}^{a'} \dfrac{da}{a}(1 + \omega)\right),
\end{eqnarray}
where we define $\xi = \dfrac{\rho}{\rho_{0}}$. Differentiating $\xi$ with respect to the scale factor, we have 
\begin{eqnarray}\label{Eq 4.34}
\dfrac{d\xi}{d a}  = \xi \left(-\dfrac{3(1 + \omega)}{a}\right).
\end{eqnarray} 
Rearranging the terms, this expression gives us 
\begin{eqnarray}\label{Eq 4.35}
w = -1 - \dfrac{a}{3\xi}\dfrac{d\xi}{d a}.
\end{eqnarray}

Using Eq.(\ref{Eq 4.20}) in terms of function $g = g(z)$, that is, $3H^2  = 8\pi G\rho g$ and $\xi = \dfrac{\rho}{\rho_{0}}$, we have
\begin{eqnarray}\label{Eq 4.36}
w = -1 -\dfrac{a}{3} \left( \dfrac{2}{H} \dfrac{d H}{d a} - \dfrac{1}{g} \dfrac{d g}{d a}\right).
\end{eqnarray}
Through the relation $a=(1+z)^{-1}$, we have
\begin{eqnarray}\label{Eq 4.37}
da = - (1+z)^{-2}dz,
\end{eqnarray}
which allows us to rewrite Eq.(\ref{Eq 4.36}) in terms of the redshift derivative. Therefore, after substituting the relation $a = (1+z)^{-1}$ and Eq.(\ref{Eq 4.37}) in Eq.(\ref{Eq 4.36}), performing the necessary operations, we obtain
\begin{eqnarray}\label{Eq 4.38}
w = -1 + \dfrac{2}{3}\dfrac{(1+z)}{H}\dfrac{d H}{dz} - \dfrac{1}{3}\dfrac{(1+z)}{g}\dfrac{d g}{dz}.
\end{eqnarray}

In the case where $w$ is constant, we obtain
\begin{eqnarray}\label{Eq 4.39}
\rho = \rho_{0}\left(\dfrac{a}{a_{0}} \right)^{-3(1+w)},
\end{eqnarray}
that is, the same result found in the standard model.

\section{Cosmological parameters}

The expression that provides us with a special value for the Universe to become spatially flat ($k=0$) for this context is obtained from the first Friedmann equation given by Eq.(\ref{Eq 4.20}). In this case, the critical density is defined by
\begin{eqnarray}\label{Eq 4.40}
\rho_{\textrm{cri}} = \dfrac{3H^2}{8\pi Gf}.
\end{eqnarray}
In order to study how the universe might evolve, we need some idea of what is in it. A more general situation is when you have a mixture of matter and radiation. Thus, we have that the total energy density is written in the form $\rho = \rho_{m} + \rho_{\gamma}$.

Here, we will express the Hubble parameter in terms of redshift and the density parameter, which is defined by $\Omega(t) = \dfrac{\rho}{\rho_{\textrm{cri}}}$. Through Eq.(\ref{Eq 4.39}) and from the relation $a=(1+z)^{-1}$ we can express $\rho_{m}$ and $\rho_{\gamma}$. For the density of matter, we have $w=0$. Thus, we obtain
\begin{eqnarray}\label{Eq 4.41}
\rho_{m} = \rho_{0,m}(1+z)^3.
\end{eqnarray}
For the case of radiation, we have $w=\frac{1}{3}$. From this, it follows that
\begin{eqnarray}\label{Eq 4.42}
\rho_{\gamma} = \rho_{0,\gamma}(1+z)^4.
\end{eqnarray}
Thus, the density parameter for each material is
\begin{eqnarray}\label{Eq 4.43}
\Omega_{0,m} = \dfrac{\rho_{0,m}}{\rho_{0,\textrm{cri}}} \ \ \ \textrm{e} \ \ \ \Omega_{0,\gamma} = \dfrac{\rho_{0,\gamma}}{\rho_{0,\textrm{cri}}}.
\end{eqnarray}
Therefore, the equation $3H^2 = 8\pi G \rho f$ is
\begin{eqnarray}\label{Eq 4.44}
3H^2 = 8\pi G (\rho_{m} + \rho_{\gamma}) f.
\end{eqnarray}
Substituting equations (\ref{Eq 4.41}), (\ref{Eq 4.42}), and (\ref{Eq 4.43}) into (\ref{Eq 4.44}), we obtain
\begin{eqnarray}\label{Eq 4.45}
3H^2 = 8\pi G \rho_{0,\textrm{cri}} \left[ \Omega_{0,m}(1+z)^3 + \Omega_{0,\gamma}(1+z)^4\right]f.
\end{eqnarray}
From Eq.(\ref{Eq 4.40}), we have that
\begin{eqnarray}\label{Eq 4.46}
\rho_{0,\textrm{cri}} = \dfrac{3H_{0}^2}{8\pi G},
\end{eqnarray}
because the value of function $f$ today is 1. Therefore, by substituting Eq.(\ref{Eq 4.46}) in Eq.(\ref{Eq 4.45}) we obtain
\begin{eqnarray}\label{Eq 4.47}
H^2 = H_{0} \left[ \Omega_{0,m}(1+z)^3 + \Omega_{0,\gamma}(1+z)^4\right]f.
\end{eqnarray}
It is interesting to note that this expression changes over time, since $H$ and $f$ also evolve.

\section{Conclusion}
This paper consists in applying convolution in the sense of Colombeau algebras, to deal with a function of distributional nature $f(t)$ capable of performing the sign change smoothly and we can work with its derivative and its powers, thus we construct the equations presented here. In our approach, we choose a way to present the function $f(t)$ that reinterprets the results from the literature. We take as a starting point the fact that the function changes sign without passing through the origin. By doing so, we avoid problems with $t = 0$ and derive the standard model equations in the context of the altered FLRW metric. With this, we can make explicit the Friedmann equations influenced by $f(t)$, show that the conservation equation is not influenced by this function, and also show the conditions for the accelerated expansion of the universe as well as the equation of state in this context. Finally, we determine the critical density and the Hubble parameter in terms of the density parameter. 

In our work, the results obtained show a physically interesting description compared to the works \cite{Mansouri1999,Nozari2006}, since the transition occurs in $t>0$, after the Big Bang ($t=0$) there is a time interval $t>0$ small enough that the Universe was ruled by a Euclidean regime, and so, as the Universe evolves space-time becomes Lorentzian and goes on as we know it today.

\end{document}